\documentclass[twocolumn,amsmath,amssymb,aps]{revtex4}
\usepackage{graphicx}
\usepackage{amsmath,amssymb}
\usepackage{mathrsfs}
\usepackage{color}
\usepackage{afterpage}
\usepackage[version=3]{mhchem}
\usepackage{natbib}
\usepackage{soul}
\usepackage[caption=false]{subfig}
\usepackage{array}
\usepackage{multirow}
\usepackage[colorlinks, citecolor=blue,urlcolor=blue, linkcolor=blue, bookmarks=false]{hyperref}
\hypersetup{colorlinks=true , citecolor=blue, urlcolor=blue, linkcolor=blue}
\usepackage{braket}
\usepackage{bm}
\begin{document}
\title{Inducing Spin Splitting and Anomalous Valley Hall Effect in A-Type AFM Fe$_2$C(OH)$_2$ through Electric Field and Janus Engineering}

\author{Ankita Phutela\footnote{ankita@physics.iitd.ac.in}, Saswata Bhattacharya\footnote{saswata@physics.iitd.ac.in}} 
\affiliation{Department of Physics, Indian Institute of Technology Delhi, New Delhi 110016, India}
\begin{abstract}
	\noindent 

The antiferromagnetic (AFM) materials are distinguished by zero net magnetic moment, high resistance to external magnetic disturbances, and ultrafast dynamic responses. For advancing AFM materials in spintronic and valleytronic applications, achieving spontaneous valley polarization and the anomalous valley Hall effect (AVHE) is pivotal. We predict an A-type AFM monolayer Fe$_2$C(OH)$_2$, which shows a significant spontaneous valley polarization of 157 meV. In Fe$_2$C(OH)$_2$, spatial inversion symmetry (\textit{P}) and time-reversal symmetry (\textit{T}) are individually broken, yet the combined \textit{PT} symmetry is preserved. This symmetry conservation leads to spin degeneracy, resulting in zero Berry curvature in the momentum space and absence of AVHE. However, a layer-locked hidden Berry curvature is produced, leading to the observation of the valley layer-spin Hall effect. Further, an external out-of-plane electric field can induce spin splitting by introducing layer-dependent electrostatic potential, enabling the layer-locked AVHE. Additionally, the introduction of a built-in electric field caused by the Janus structure also induces spin splitting in monolayer Fe$_2$C(OH)F due to the electric-potential-difference-AFM mechanism. The high out-of-plane magnetic anisotropy and realization of AVHE, offer promising opportunities for next-generation spintronic technologies.

\end{abstract}
\maketitle

\section{Introduction}
In two-dimensional (2D) materials, valleys represent the local energy extrema in the conduction and valence bands, serving as an additional degree of freedom alongside charge and spin ~\cite{rycerz2007valley,shkolnikov2002valley}. Valleytronics is an emerging field that exploits this valley degree of freedom by manipulating the population of electrons in different valleys ~\cite{xiao2007valley,schaibley2016valleytronics,gunawan2006valley,xiao2012coupled,rycerz2007valley}. The ability to selectively populate one valley over another is known as valley polarization, and is essential for using the valleys as information carriers. This polarization creates an imbalance between the valleys, allowing for the development of low-power, high-speed information processing, and logic devices ~\cite{sahoo2022enhanced,pawlowski2021valley,chen2022room,pacchioni2020valleytronics}. For nonmagnetic materials, 2H-transition metal dichalcogenides have attracted
extensive attention as they have a pair of degenerate valleys at a large intervalley distance ~\cite{sheoran2023probing,sheoran2023coupled,srivastava2015valley,yao2008valley,macneill2015breaking,zhao2017enhanced,zeng2018exploring}. They show opposite spin splitting at $-$K and $+$K valleys, which can be called spin-valley locking, but lack spontaneous valley polarization hindering further extension of valleytronics. To manipulate valley polarization, various external methods such as magnetic fields ~\cite{macneill2015breaking}, optical excitation, ~\cite{srivastava2015valley,yao2008valley} and magnetic doping ~\cite{zeng2018exploring} have been employed, but they bring many undesirable effects ~\cite{macneill2015breaking,stier2016exciton}.

In most recent works, spontaneous valley polarization has been predicted in 2D hexagonal ferromagnetic (FM) materials such as VSi$_2$P$_4$ ~\cite{feng2021valley}, Cr$_2$Se$_2$ ~\cite{he2021two}, Cr$_2$S$_2$ ~\cite{li2023spontaneous}, and Cr$_2$CSF ~\cite{phutela2024probing}. Compared to these FM materials, the antiferromagnetic (AFM) materials provide an intriguing alternative for valleytronic applications ~\cite{guo2024electric,guo2024large,li2023multifield}. They possess zero magnetic moments, are inherently robust to external magnetic perturbations, and possess ultrafast dynamics, thus providing enormous potential for device applications. They have inherent valley polarization but these antiferromagnets generally lack spontaneous spin splitting in the band structures, which prevents some interesting physical phenomena ~\cite{zhang2024spontaneous}. AFM materials with intrinsic spin splitting are particularly important, as they allow for the realization of anomalous valley Hall effect (AVHE) without the need for external manipulation. However, in AFM materials with combined \textit{PT} symmetry (where \textit{P} represents spatial inversion symmetry and \textit{T} is time-reversal symmetry), there is zero Berry curvature and no spin splitting, which hinders the realization of AVHE ~\cite{guo2024large}. This spin degeneracy can be lifted by the application of an external electric field or by altering atomic arrangements, resulting in observable spin splitting in the electronic band structure ~\cite{guo2024layer}. The atomic alteration can be achieved by placing magnetic atoms with opposite spin polarizations in different environments. When these atoms experience different surrounding atomic arrangements, the symmetry that protects spin degeneracy is broken. This mechanism is known as electric-potential-difference antiferromagnetism (EPD-AFM) ~\cite{guo2023spontaneous}. 

In this work, we have identified Fe$_2$C(OH)$_2$ as a promising A-type AFM valleytronic material exhibiting high spontaneous valley polarization. Fe$_2$C(OH)$_2$ lacks intrinsic spin splitting in its band structure due to conservation of combined \textit{PT} symmetry, resulting in zero Berry curvature in the momentum space. This prevents AVHE, however, a valley layer-spin Hall effect is observed. The spin splitting is induced through the application of an external out-of-plane electric field. By combining with layer-locked Berry curvature, the layer-locked AVHE can be achieved in monolayer Fe$_2$C(OH)$_2$ under the external electric field. Additionally, Janus Fe$_2$C(OH)F is constructed to achieve EPD-AFM, which is a promising candidate material for realizing AVHE. A high out-of-plane magnetic anisotropy energy (MAE) depicted by both Fe$_2$C(OH)$_2$ and Fe$_2$C(OH)F makes them desirable for valleytronic device application.
 
\section{Computational Methods}
The first-principles calculations based on the density functional theory (DFT) are carried out ~\cite{hohenberg1964inhomogeneous,kohn1965self} using the Vienna \textit{ab initio} Simulation Package (VASP) ~\cite{kresse1996efficient} with the projector augmented wave (PAW) ~\cite{kresse1999ultrasoft,blochl1994projector} method. The generalized gradient approximation (GGA) is used to describe the electron exchange-correlation interactions in the form of Perdew-Burke-Ernzerhof (PBE) functional ~\cite{perdew1996generalized}. The cutoff energy for the plane wave basis is set to 520 eV. All the structures are optimized until the Hellmann-Feynman force on each ion is less than 1 meV/\AA, and the self-consistent convergence criterion of the electronic iteration is set to 10$^{-5}$ eV. The vacuum space in \textit{z}-direction is set to 30 \AA\ to avoid the interactions between the periodically adjacent layers. The 11$\times$11$\times$1 \textit{k}-grid is used to sample the 2D Brillouin zone. To account for the localized nature of Fe-3\textit{d} orbitals, a Hubbard correction is used by the rotationally invariant approach proposed by Dudarev et al. ~\cite{dudarev1998electron}. The linear response method is used to determine the effective U (U$_{eff}$) values, and they are 5.3 and 5.5 for Fe$_2$C(OH)$_2$ and Fe$_2$C(OH)F, respectively ~\cite{cococcioni2005linear}. In addition, hybrid functional Heyd-Scuseria Ernzerhof (HSE06) is also employed to confirm the results of GGA+U calculations ~\cite{heyd2003hybrid}. The spin-orbit coupling (SOC) effect is considered in the calculations to address the relativistic effect, as implemented in VASP by noncollinear calculations. The Berry curvatures are computed directly from the calculated wave functions using Fukui's method, as implemented in the VASPBERRY code ~\cite{kim2016competing}.

\section{RESULTS AND DISCUSSION}

\begin{figure}[htp]
	\includegraphics[width=0.5\textwidth]{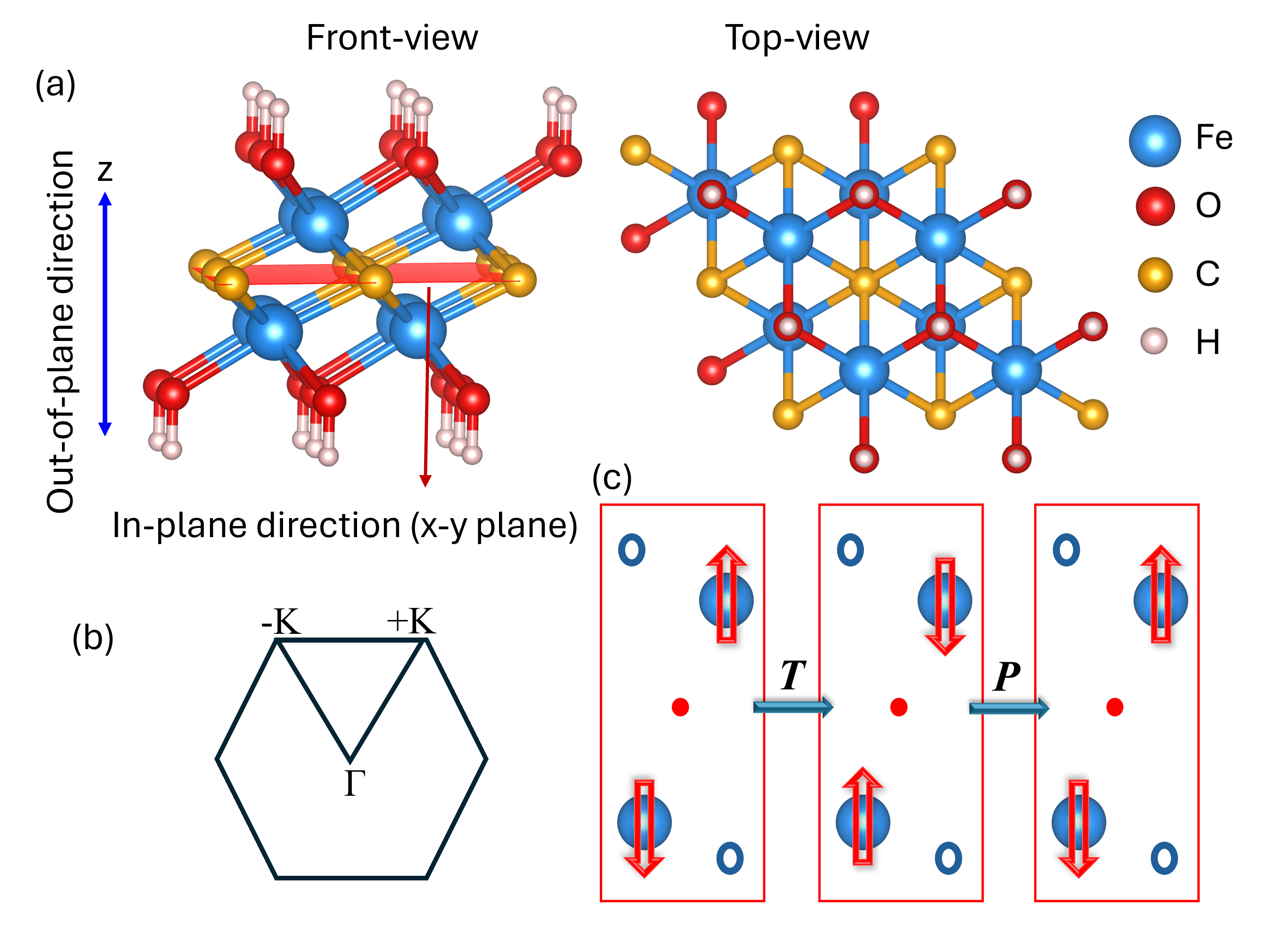}
	\caption{(a) Front and top views of Fe$_2$C(OH)$_2$ (Here, \(x\)-\(y\) plane represents the in-plane direction and the \(z\)-axis represents the out-of-plane direction). (b) Hexagonal 2D Brillouin zone. (c) A combination of \textit{P} and \textit{T} symmetry showing combined \textit{PT} symmetry.}
	\label{pic1}
\end{figure}

Fe$_2$C-based MXenes have been reported to exhibit high Curie/N\'eel temperatures and dynamic stabilities ~\cite{hu2020high}. We investigate the valley properties and band structures of Fe$_2$C-based MXenes: Fe$_2$C(OH)$_2$, Fe$_2$CF$_2$, Fe$_2$CCl$_2$, Fe$_2$CO$_2$, and Fe$_2$CS$_2$. Fe$_2$C(OH)$_2$ is explored as a prototype material in our study due to its promising valleytronic properties, which will be discussed comprehensively in subsequent sections. The crystal structure of Fe$_2$C(OH)$_2$, along with the corresponding 2D Brillouin zone, is shown in Fig. \ref{pic1}(a) and (b), respectively. Fe$_2$C(OH)$_2$ consists of seven atomic layers arranged in the sequence H-O-Fe-C-Fe-O-H, with the central carbon layer sandwiched between two Fe-OH bilayers. This compound crystallizes in \textit{P}3\textit{m}1 space group (No. 164), exhibiting spatial inversion symmetry in the absence of magnetic ordering. The four possible magnetic configurations namely FM, AFM1, AFM2, and AFM3 are considered as seen in section I of Supplemental Material (SM). The blue and gray spheres represent spin-up and spin-down Fe atoms, respectively. The magnetic ground state of Fe$_2$C(OH)$_2$ is determined by comparing the energies of these magnetic configurations. Among these, AFM1 is identified as the A-type antiferromagnetic state characterized by intralayer FM and interlayer AFM couplings. The energy of AFM1 configuration is 800 meV, 700 meV, and 250 meV lower than the FM, AFM2, and AFM3 configurations, respectively, confirming that  Fe$_2$C(OH)$_2$ prefers A-type AFM ordering. This has also been confirmed by using more accurate HSE06 functional (see section I of SM). The total magnetic moment per unit cell is 0.00 \( \mu_B \) (where \( \mu_B \) is Bohr magneton), with the magnetic moments of the top and bottom Fe layers being $+$3.91 \( \mu_B \) and $-$3.91 \( \mu_B \), respectively. In this A-type AFM state, the spin configuration breaks the \textit{P} and \textit{T} symmetry. However, the combined \textit{PT} symmetry remains intact in Fe$_2$C(OH)$_2$ as shown in Fig. \ref{pic1}(c). The \textit{T} operation reverses the spin direction of each Fe atom, followed by the \textit{P} operation which swaps the top and bottom Fe atoms, resulting in symmetry conservation.

\begin{figure}[htp]
	\includegraphics[width=0.45\textwidth]{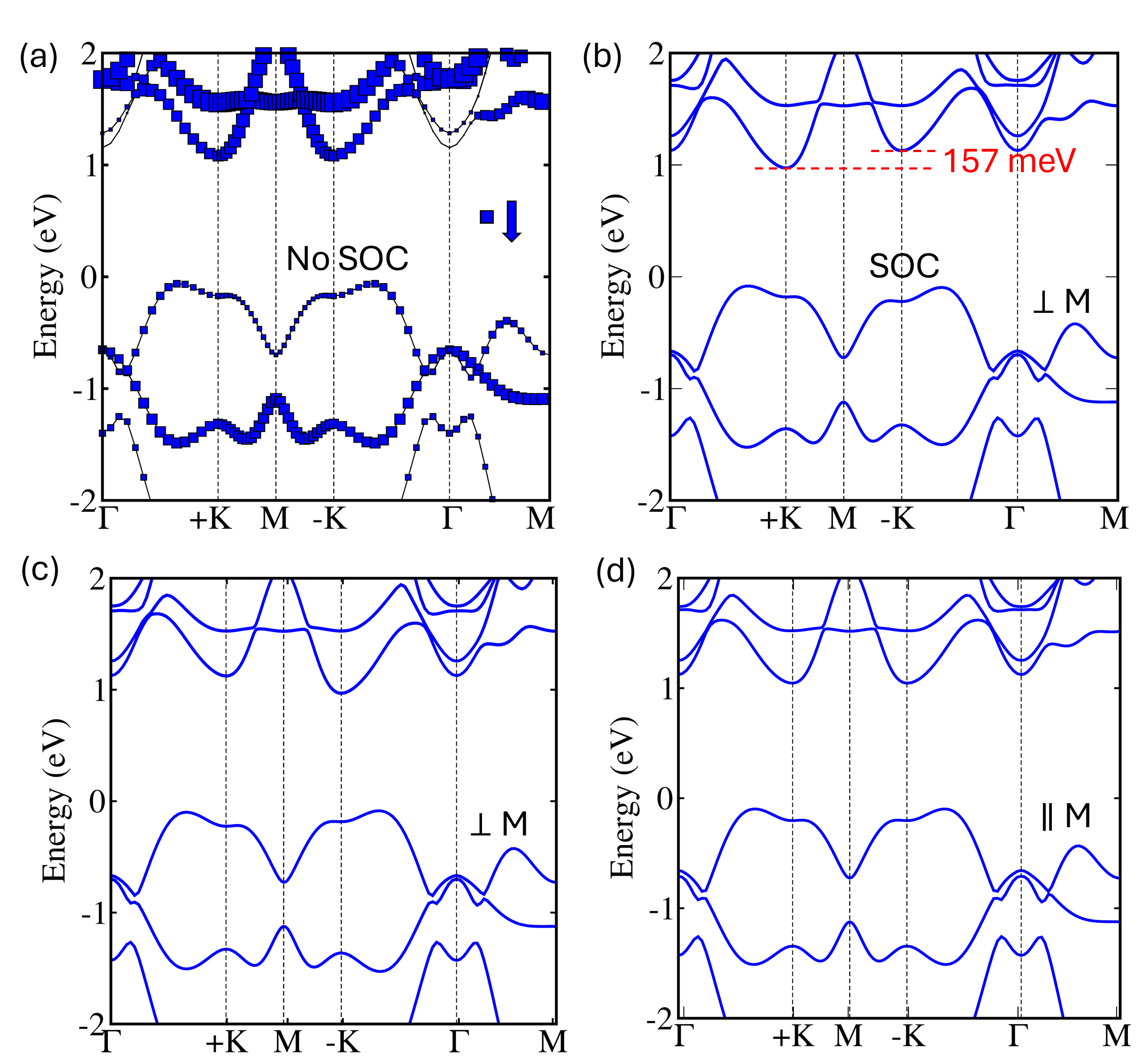}
	\caption{The energy band structures of Fe$_2$C(OH)$_2$ (a) without SOC with blue colour showing the contribution of spin-down states only,  with SOC (b) for out-of-plane magnetization (i.e. along the positive \(z\)-direction) showing valley polarization, (c) with opposite out-of-plane magnetization (along the negative \(z\)-direction), and (d) for in-plane magnetization, showing absence of valley polarization.}
	\label{pic2}
\end{figure}
 
The electronic band structures for Fe$_2$CF$_2$, Fe$_2$CCl$_2$, Fe$_2$CO$_2$, and Fe$_2$CS$_2$ are displayed in section II of SM and that of Fe$_2$C(OH)$_2$ is shown in Fig. \ref{pic2}(a). The energy band structure shows that Fe$_2$C(OH)$_2$ is an indirect band-gap semiconductor with degenerate energy at $+$K and $-$K points in the conduction band. The valleys are observed at $+$K and $-$K points, but no spin splitting is observed due to the preservation of \textit{PT} symmetry. As from \textit{P} symmetry: $E(\mathbf{k}, \uparrow) = E(-\mathbf{k}, \uparrow)$, from \textit{T} symmetry: $E(-\mathbf{k}, \uparrow) = E(\mathbf{k}, \downarrow)$, and combining these two leads to \textit{PT} symmetry: $E(\mathbf{k}, \uparrow) = E(\mathbf{k}, \downarrow)$. This means that the spin-up and spin-down bands are degenerate at every \textbf{k}, and there is no spin splitting in the band structure. When SOC is considered, as shown in Fig. \ref{pic2}(b), the energy at $-$K valley becomes higher than that of $+$K valley. The opposite spin vectors of the two sublattices in Fe$_2$C(OH)$_2$ break the \textit{P} and \textit{T} symmetry giving rise to spontaneous valley polarization. A notable valley polarization is induced with a sizable valley splitting of 157 meV ($\Delta$$E_C$ =$ E_{-K}$ $-$ $E_{+K}$). This magnitude of valley splitting is higher than many reported magnetic valley materials ~\cite{tong2016concepts,du2022anomalous,zhao2024stacking,guo2024large}. Further, when the direction of magnetization is reversed, the valley polarization gets switched with the energy at $+$K valley becoming higher than that of $-$K valley with the same magnitude of valley splitting (see Fig. \ref{pic2}(c)). When the magnetization direction is in-plane, no valley polarization is observed as shown in Fig. \ref{pic2}(d). Even with SOC included, the spin degeneracy is maintained for both in-plane and out-of-plane magnetization directions. This indicates that Fe$_2$C(OH)$_2$ exhibits robust valleytronic behavior. Furthermore, the MAE, which determines the preferred magnetization direction, is calculated using the equation: MAE = $E_{x/y}$ $-$ $E_z$, where $E_{x/y}$ and $E_z$  denote the in-plane and out-of-plane spin orientations, respectively. The MAE value is 174 $\mu$eV/unit cell, confirming that Fe$_2$C(OH)$_2$ has an out-of-plane easy magnetization axis. This out-of-plane magnetization leads to spontaneous valley polarization, supporting the realization of valleytronic applications. 
\begin{figure}[htp]
	\includegraphics[width=0.5\textwidth]{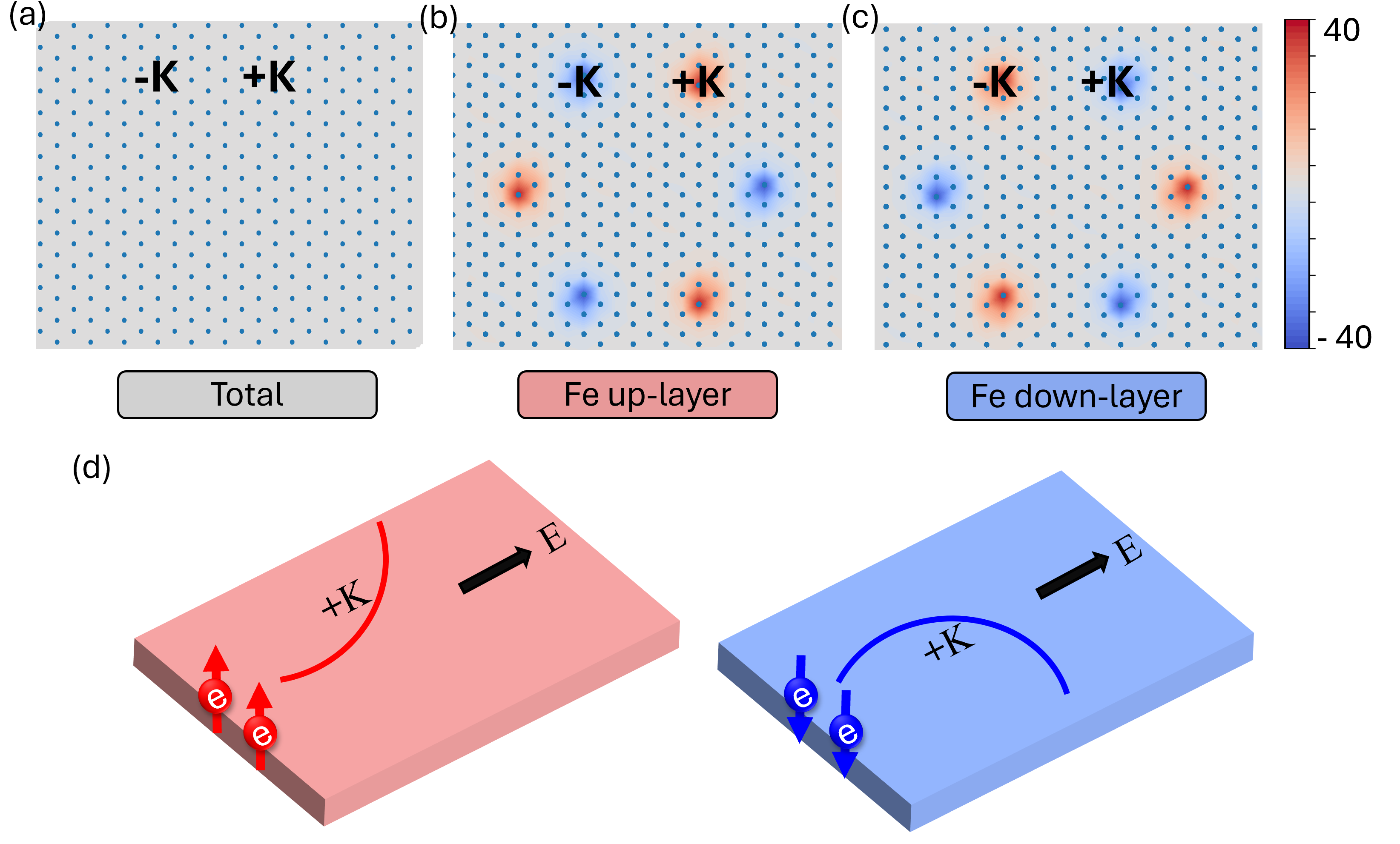}
	\caption{The distribution of (a) total, (b) spin-up, and (c) spin-down Berry curvatures for Fe$_2$C(OH)$_2$. (d) A schematic of valley layer-spin Hall effect under the application of a longitudinal in-plane electric field and appropriate electron doping. The left and right planes represent the top and bottom Fe layers, respectively.}
	\label{pic3}
\end{figure}
The SOC-induced valley polarization primarily arises from the intra-atomic interaction $\hat{H}_{\text{SOC}}^0$, which affects the states with the same spin orientation and is given by: ~\cite{wang1996torque,dai2008effects}.

\begin{equation}
\hat{H}_{\text{SOC}}^0 = \lambda \hat{S}_{z^\prime} \left (\hat{L}_z \cos\theta + \frac{1}{2} \hat{L}_+ e^{-i\phi} \sin\theta + \frac{1}{2} \hat{L}_- e^{i\phi} \sin\theta\right)
\end{equation}
\noindent Here, the magnetization is out-of-plane, i.e., \(\theta\) = 0 (see section III of SM). Therefore, the above equation is simplified to:

\begin{equation}
	\hat{H}_{\text{SOC}}^0 = \lambda \hat{S}_{z^\prime} \hat{L}_z 
\end{equation}
\noindent To proceed further, we have observed that the conduction band valleys are primarily composed of $d_{x^2-y^2}$ and $d_{xy}$ orbitals of the Fe atoms (see section III of SM). Therefore, the basis wave vectors are chosen in the following form:
\begin{equation}
|\phi_{c}^\tau\rangle = \frac{1}{\sqrt{2}} \left(|d_{x^2-y^2}\rangle + i\tau|d_{xy}\rangle \right) \
\end{equation}
where c represents the conduction band and $\tau$ = $\pm$1 corresponds to $+$K/$-$K valley. The energy level at $+$K/$-$K can be denoted by following formula:
\begin{equation}
\begin{aligned}
E_{c}^{+\mathrm{K}/-\mathrm{K}} &= \langle\phi_{c}^\tau|\hat{H}_{\text{SOC}}^0|\phi_{c}^\tau\rangle \\
& = \frac{i\tau}{2}(\langle d_{x^2-y^2}|\hat{H}_{\text{SOC}}^0|d_{xy}\rangle\\ & - \langle d_{xy}|\hat{H}_{\text{SOC}}^0|d_{x^2-y^2}\rangle)\\
& = \frac{i\tau}{2}\lambda(\langle d_{x^2-y^2}|\hat{L}_z|d_{xy}\rangle\ -\langle d_{xy}|\hat{L}_z |d_{x^2-y^2}\rangle)
\end{aligned}
\end{equation}
where  \(\hat{L_z}\)$\ket{d_{x^2-y^2}}$ = 2$\textit{i}$$\ket{d_{xy}}$, \(\hat{L_z}\)$\ket{d_{xy}}$ = --2$\textit{i}$$\ket{d_{x^2-y^2}}$ and $\alpha$ = $\braket{\uparrow|\hat S_z|\uparrow}$. The valley polarization in the conduction band is obtained through

\begin{equation}
\begin{aligned}
\Delta E_{c} & = E_{c}^{+\mathrm{K}} - E_{c}^{-\mathrm{K}} \\
& = i\lambda\alpha\langle d_{x^2-y^2}|\hat{L}_z|d_{xy}\rangle - i\lambda\langle d_{xy}|\hat{L}_z|d_{x^2-y^2}\rangle\\
& = 4\lambda\alpha
\end{aligned}
\end{equation}

When the magnetocrystalline direction is along an out-of-plane, the valley splitting of Fe$_2$C(OH)$_2$ will be 4$\lambda\alpha$.

\noindent On the other hand, if magnetic orientation is in-plane, i.e., $\theta = \frac{\pi}{2}$, we get:

\begin{equation}
\hat{H}_{\text{SOC}}^0 = \lambda \hat{S}_{z^\prime} \left ( \frac{1}{2} \hat{L}_+ e^{-i\phi}  + \frac{1}{2} \hat{L}_- e^{i\phi} \right)
\end{equation}

\noindent In this case, \textit{$\Delta$\textit{E}$_{c}$} = 0, i.e., no spontaneous valley polarization (Fig. \ref{pic2}(d)).

\begin{figure}[h]
	\includegraphics[width=0.4\textwidth]{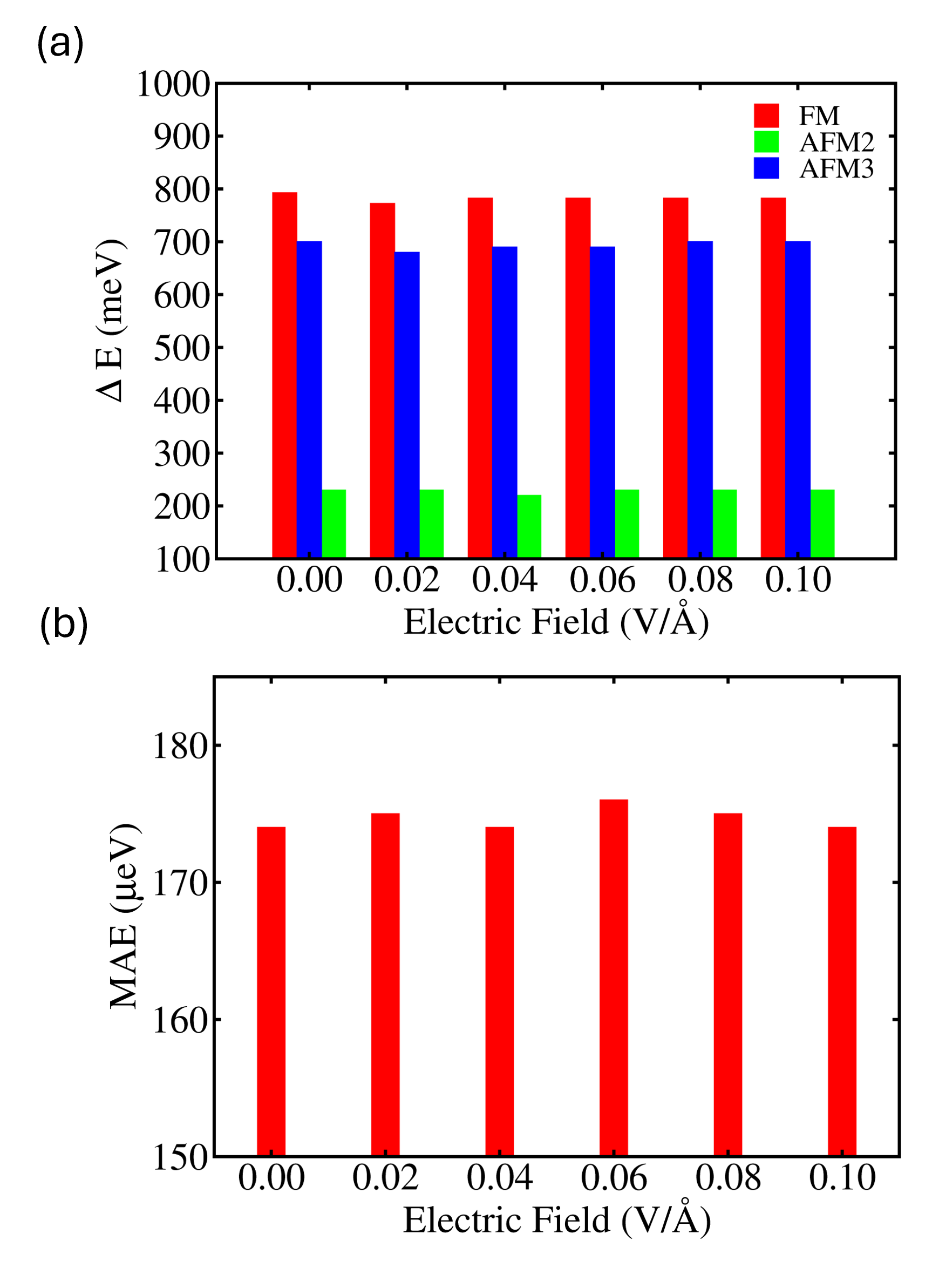}
	\caption{For Fe$_2$C(OH)$_2$, (a) the energy difference between FM/AFM2/AFM3 and AFM1 magnetic orderings as a function of electric field, and (b) the variation of MAE as a function of electric field.}
	\label{pic4}
\end{figure}

The distribution of total Berry curvature for Fe$_2$C(OH)$_2$ is shown in Fig. \ref{pic3}(a). Due to the conserved \textit{PT} symmetry, the total Berry curvature is zero throughout the momentum space. Therefore, Fe$_2$C(OH)$_2$ cannot show AVHE but it can show valley layer-spin Hall effect. Figure \ref{pic3}(b) and (c) show the Berry curvatures of spin-up and spin-down channels. The Berry curvatures show opposite signs for different valleys of the same spin channel and the same valley of different spin channels, producing a layer-locked hidden Berry curvature. This Berry curvature arises because each layer unit (C-Fe-OH) breaks the local \textit{PT} symmetry. When the Fermi level is shifted between the $+$K and $-$K valleys in the conduction band via electron doping, the spin-up and spin-down electrons from $+$K valley will accumulate along the opposite sides of the specimen under the effect of an in-plane electric field. Under this field, the Bloch electrons will gain an anomalous velocity, equal to the cross product of in-plane electric field and Berry curvature i.e., \(\upsilon \sim E \times \Omega(\mathbf{k})\) ~\cite{sheoran2023manipulation,xu2014spin,xiao2010berry}. This results in the valley layer-spin Hall effect (Fig. \ref{pic3}(d)).

\begin{figure}[h]
	\includegraphics[width=0.5\textwidth]{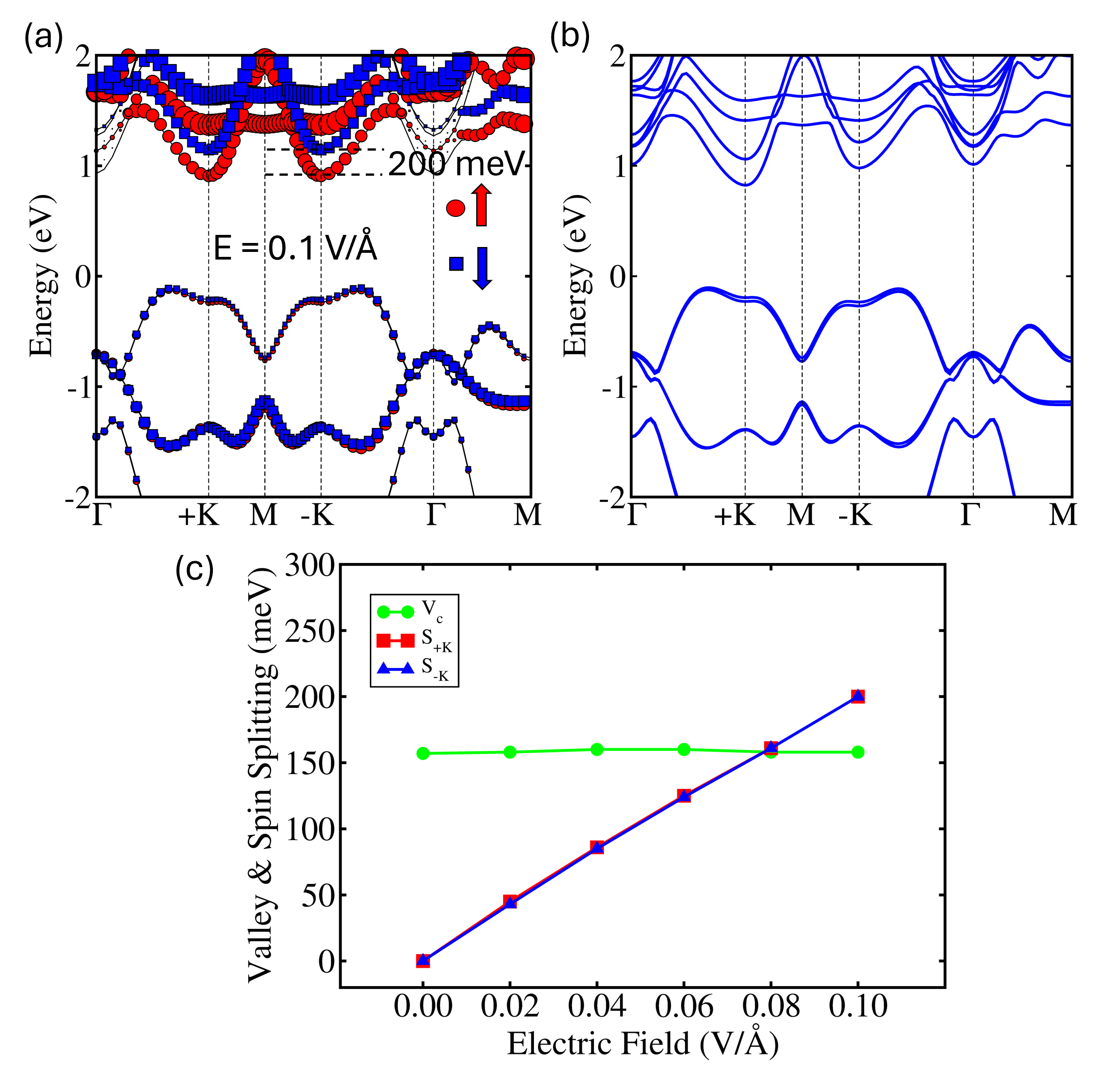}
	\caption{For Fe$_2$C(OH)$_2$, the spin-resolved electronic band structure at representative electric field: E = 0.10 V/\AA\ (a) in absence of SOC showing spin splitting, and (b) in presence of SOC, showing spin split bands along with valley polarization. (c) The magnitude of valley splitting (V$_c$), and spin splitting (S$_{+K}$ and S$_{-K}$ at $+$K and $-$K valleys, respectively) in the conduction band as a function of electric field.}
	\label{pic5}
\end{figure}

Further, an out-of-plane electric field is employed to break the \textit{PT} symmetry ~\cite{guo2024electric}. This removal subsequently leads to the lifting of spin degeneracy in the band structure. A layer-dependent electrostatic potential induced by the out-of-plane electric field results in the spin splitting effect. An out-of-plane electric field ranging from 0.00 to 0.10 V/\AA\ is applied. Firstly, we have determined the magnetic ground state by calculating the energy of FM, AFM1, AFM2, and AFM3 configurations under the applied range of electric field. As shown in Fig. \ref{pic4}(a), AFM1 state remains the preferred magnetic ground state at all the electric fields and the energy does not vary much over this range. Subsequently, we have calculated the MAE vs. electric field and concluded that the MAE consistently favors the out-of-plane direction and shows minimal variation under the applied electric field (Fig. \ref{pic4}(b)). The electronic band structures, without and with SOC are computed for various electric field values (0.02, 0.04, 0.06, 0.08 V/\AA) as shown in section IV of SM. Figure \ref{pic5}(a) shows the energy band structure at a representative electric field 0.10 V/\AA\ without including SOC. It is evident from the plots that spin splitting begins to occur upon the application of electric field. When an electric field of 0.02 V/\AA\ is applied a spin splitting of magnitude of 45 meV is observed. Theoretically, the magnitude of spin splitting $\propto$ eEd, where E is the applied electric field and d is the separation over which the potential difference acts ~\cite{guo2024electric}. For Fe$_2$C(OH)$_2$, d = 2.55 \AA, and the calculated splitting is 51 meV, which is close to our first-principles value. At a representative value of E = 0.10 V/\AA\ the spin splitting of magnitude 200 meV is observed as shown in Fig. \ref{pic5}(a). When SOC is applied, the valley polarization appears along with the spin splitting as shown in Fig. \ref{pic5}(b). As seen from Fig. \ref{pic5} (c), the magnitude of valley polarization is least impacted by the magnitude of electric field. Although the magnitude of spin splitting as expected varies linearly with the electric field. Moreover, the spin splitting has the same value at $+$K and $-$K valleys. The coexistence of valley polarization and spin splitting is needed for the realization of AVHE. When the direction of electric field is reversed from $+$E to $-$E, the layer-dependent electrostatic potential is also reversed, reversing the spin order of spin splitting (see section V of SM). However, the size of spin splitting and valley polarization remain unchanged. This happens due to the relationship between two Fe layers, which possess opposite magnetic moments and are connected by a glide mirror symmetry G$_z$.

 \begin{figure}[htp]
 	\includegraphics[width=0.45\textwidth]{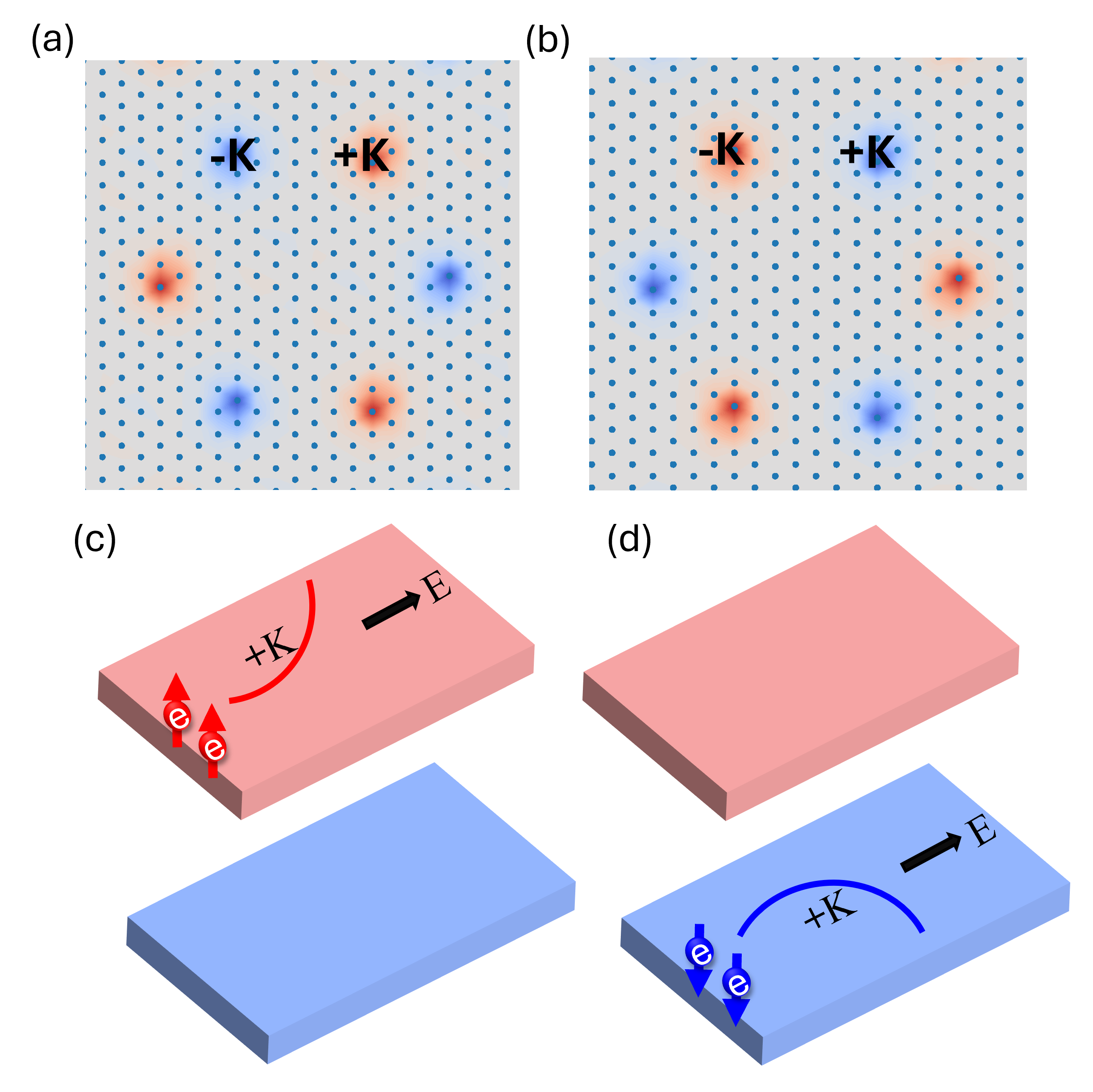}
 	\caption{The distribution of (a) spin-up, and (b) spin-down Berry curvatures for  Fe$_2$C(OH)$_2$ under an electric field of 0.10 V/\AA. (c), (d) The in-plane electric field and electron doping produce layer-locked AVHE. The top and bottom planes represent the top and bottom Fe layers, respectively.}
 	\label{pic6}
 \end{figure}
 
Further, the distribution of Berry curvatures for up, and down spins under the electric field of magnitude 0.10 V/\AA\ are plotted in Fig. \ref{pic6}(a) and (b). It is seen that the Berry curvature is opposite at $+$K and $-$K points. Upon shifting the Fermi level between $+$K and $-$K valley of the conduction band and application of an in-plane electric field, the spin-up electrons in the conduction band at $+$K move under the action of an anomalous velocity. The electrons get concentrated at the top edge of the sample as illustrated in Fig. \ref{pic6}(c). This leads to a layer-locked AVHE. When the direction of electric field is reversed, the valleys in the conduction band are formed of spin-down electrons. Under the action of negative Berry curvature and an in-plane electric field, the spin-down electrons at $+$K move to the lower edge of the sample. This accumulation of spin-polarized electrons produces a net spin current, and a voltage is generated.

\begin{figure}[htp]
	\includegraphics[width=0.5\textwidth]{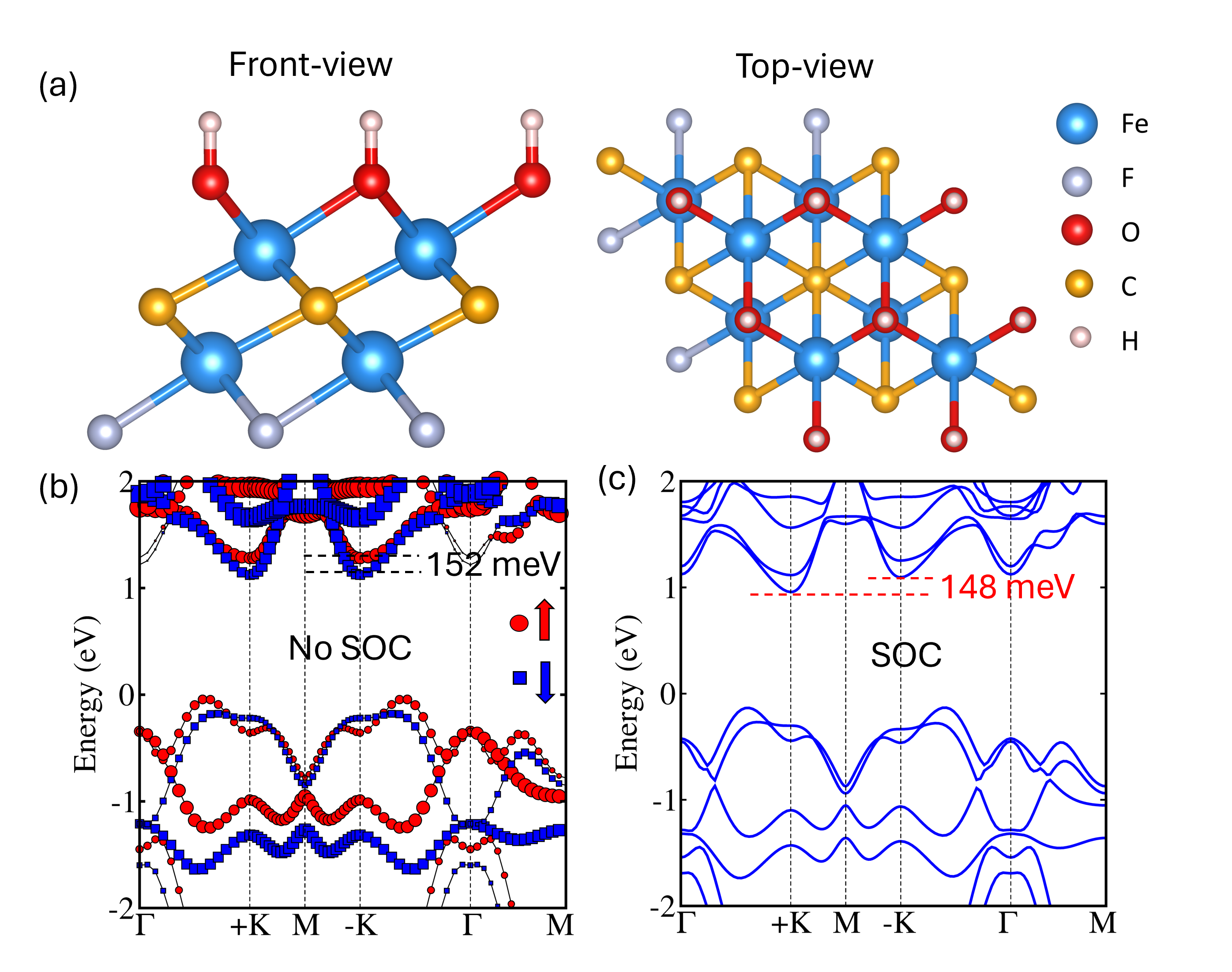}
	\caption{(a) The top and (b) side views of the crystal structure of Janus monolayer Fe$_2$C(OH)F. The energy band structure (c) without	SOC with red (blue) circles (squares) representing the spin-up (spin-down) states, and (d) with SOC for magnetization direction along the positive \(z\)-direction.}
	\label{pic7}
\end{figure}
The two Fe atoms in Fe$_2$C(OH)$_2$ have the same surrounding atomic environment of C and OH, which gives rise to degeneration of electron spin. To break this degeneracy, a Janus monolayer can be constructed by replacing the OH layer with the F element, forming Fe$_2$C(OH)F. The front and top views of Fe$_2$C(OH)F are shown in Fig. \ref{pic7}(a). The structure consists of six atomic layers arranged in the sequence H-O-Fe-C-Fe-F, where each Fe atom now experiences a different surrounding atomic environment. The difference in electronegativity of OH and F induces an intrinsic polar electric field along the \(z\)-direction, enabling the realization of EPD-AFM. The energy of FM and three AFM configurations (AFM1, AFM2 and AFM3) is calculated to determine the magnetic ground state of Fe$_2$C(OH)F. The AFM1 ordering is found to be the magnetic ground state with energies 1.02 eV, 232 meV, and 713 meV less than FM, AFM2, and AFM3 configurations, respectively. The total magnetic moment of Fe$_2$C(OH)F per unit cell is 0.00 $\mu_B$, with individual magnetic moments of 3.90 $\mu_B$ and $-$3.85 $\mu_B$ for the top and bottom Fe layers, respectively. To confirm the spontaneous valley polarization of Fe$_2$C(OH)F, the MAE is calculated, and the predicted value is 169 $\mu$eV/unit cell, which means that the easy magnetization axis is out-of-plane. As seen in  Fig. \ref{pic7}(b), due to the broken \textit{PT} symmetry, the spin splitting of magnitude 152 meV is observed in the energy band structure. When SOC is included with the magnetization direction along positive \(z\)-direction, the spontaneous valley polarization of magnitude 148 meV appears (Fig. \ref{pic7}(c)). The Berry curvatures for up and down spins are calculated as seen in section VI of SM. The Berry curvatures are peaked at $+$K and $-$K valleys. It is seen that the Berry curvatures are opposite for  $+$K and $-$K valleys. Moreover, the Berry curvatures at the same valley are also opposite for different spin channels. AVHE can be observed in Fe$_2$C(OH)F under the effect of an in-plane electric field. By shifting the Fermi level between the $+$K and $-$K valleys in the conduction band, the spin-down carriers from the $+$K valley will accumulate along the bottom boundary of the sample, resulting in the layer-locked AVHE.

\section{CONCLUSIONS}
To summarize, using first-principles calculations, we have studied the valley-dependent properties of Fe$_2$C(OH)$_2$. It is an A-type AFM monolayer with a large spontaneous valley polarization of 157 meV in the conduction band. The combined \textit{PT} symmetry in Fe$_2$C(OH)$_2$ ensures spin degeneracy, resulting in zero Berry curvature in the momentum space, and consequently suppressing the AVHE. However, a nonzero layer-locked hidden Berry curvature emerges in real space, leading to the observation of valley layer-spin Hall effect. Further, by applying an out-of-plane electric field, a layer-dependent electrostatic potential is introduced, inducing spin splitting and enabling a layer-locked AVHE in Fe$_2$C(OH)$_2$. Notably, the order of spin splitting can be reversed by changing the direction of the applied electric field. Furthermore, AVHE can be intrinsically realized in Janus Fe$_2$C(OH)F without the need for an external electric field, owing to its built-in electric potential. These findings demonstrate that AFM monolayers Fe$_2$C(OH)$_2$ and Fe$_2$C(OH)F, possess exceptional properties for achieving AVHE, offering significant potential for spintronic and valleytronic applications.

\section*{ACKNOWLEDGEMENT}
A.P. acknowledges IIT Delhi for the senior research fellowship. S.B. acknowledges financial support from SERB under a core research grant (grant no. CRG/2019/000647) to set up his High Performance Computing (HPC) facility ‘‘\textit{Veena}’’ at IIT Delhi for computational resources.

	

\bibliography{bibliography.bib}
\end{document}


\begin{flushleft}	
\title{Inducing Spin Splitting and Anomalous Valley Hall Effect in A-Type AFM Fe$_2$C(OH)$_2$ through Electric Field and Janus Engineering }
\author{Ankita Phutela\footnote{Ankita@physics.iitd.ac.in}, Saswata Bhattacharya\footnote{saswata@physics.iitd.ac.in}} 
\affiliation{Department of Physics, Indian Institute of Technology Delhi, New Delhi 110016, India}
\keywords{DFT}
\maketitle
\begin{center}
	{\Large \bf Supplemental Material}\\ 
\end{center}
\begin{enumerate}[\bf I.]
	\item Crystal structures and energies of various magnetic configurations of Fe$_2$C(OH)$_2$ MXene.
	\item The electronic band structures of Fe$_2$CF$_2$, Fe$_2$CCl$_2$, Fe$_2$CO$_2$, and Fe$_2$CS$_2$.
	\item Orbital projected band structure of Fe$_2$C(OH)$_2$ MXene.
	\item The energy band structures for Fe$_2$C(OH)$_2$ at various	representative electric fields.
	\item Reversal of the spin order of spin splitting on reversing the direction of electric field.
	\item Distribution of Berry curvature for Janus Fe$_2$C(OH)F.

\end{enumerate}
\vspace*{12pt}
\clearpage

\newpage

\section{C\MakeLowercase{rystal structures and energies of various magnetic configurations of} F\MakeLowercase{e}$_2$C(OH)$_2$ MX\MakeLowercase{ene}.}
\begin{figure*}[htp]
	\includegraphics[width=0.7\textwidth]{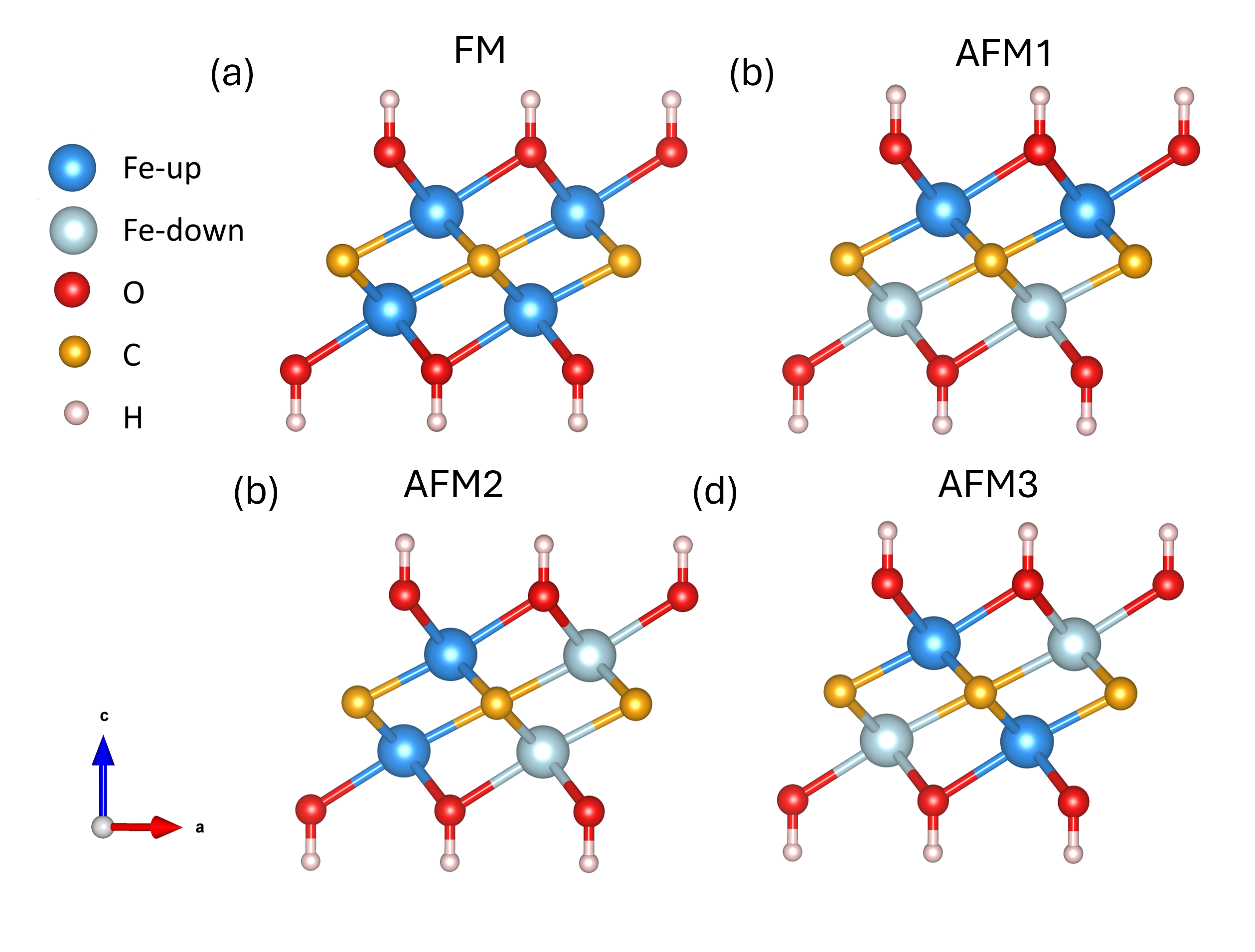}
	\caption{Crystal structures of (a) FM, (b) AFM1, (c) AFM2, and (d) AFM3 magnetic configurations for Fe$_2$C(OH)$_2$  MXene.}
	\label{p1}
\end{figure*}

\begin{table*}[htbp]
	\caption {Energy of different magnetic configurations of Fe$_2$C(OH)$_2$ MXene using HSE06 exchange functional. The energy of most stable configuration has been set to zero (in units of meV).}
	\begin{ruledtabular}	
		\begin{tabular}[c]{cccccccccccc} 		
			Fe$_2$C(OH)$_2$ MXene&  HSE06    \\ \hline
			FM  &   100 \\ 
			AFM1  & 0   \\ 
			AFM2  & 620   \\
			AFM3  &  370 \\
		\end{tabular}
	\end{ruledtabular}
	\label{T3}
\end{table*}

\newpage
\section{T\MakeLowercase{he electronic band structures of} F\MakeLowercase{e}$_2$CF$_2$, F\MakeLowercase{e}$_2$CC\MakeLowercase{l}$_2$, F\MakeLowercase{e}$_2$CO$_2$, \MakeLowercase{and} F\MakeLowercase{e}$_2$CS$_2$.}
\begin{figure*}[htp]
	\includegraphics[width=0.7\textwidth]{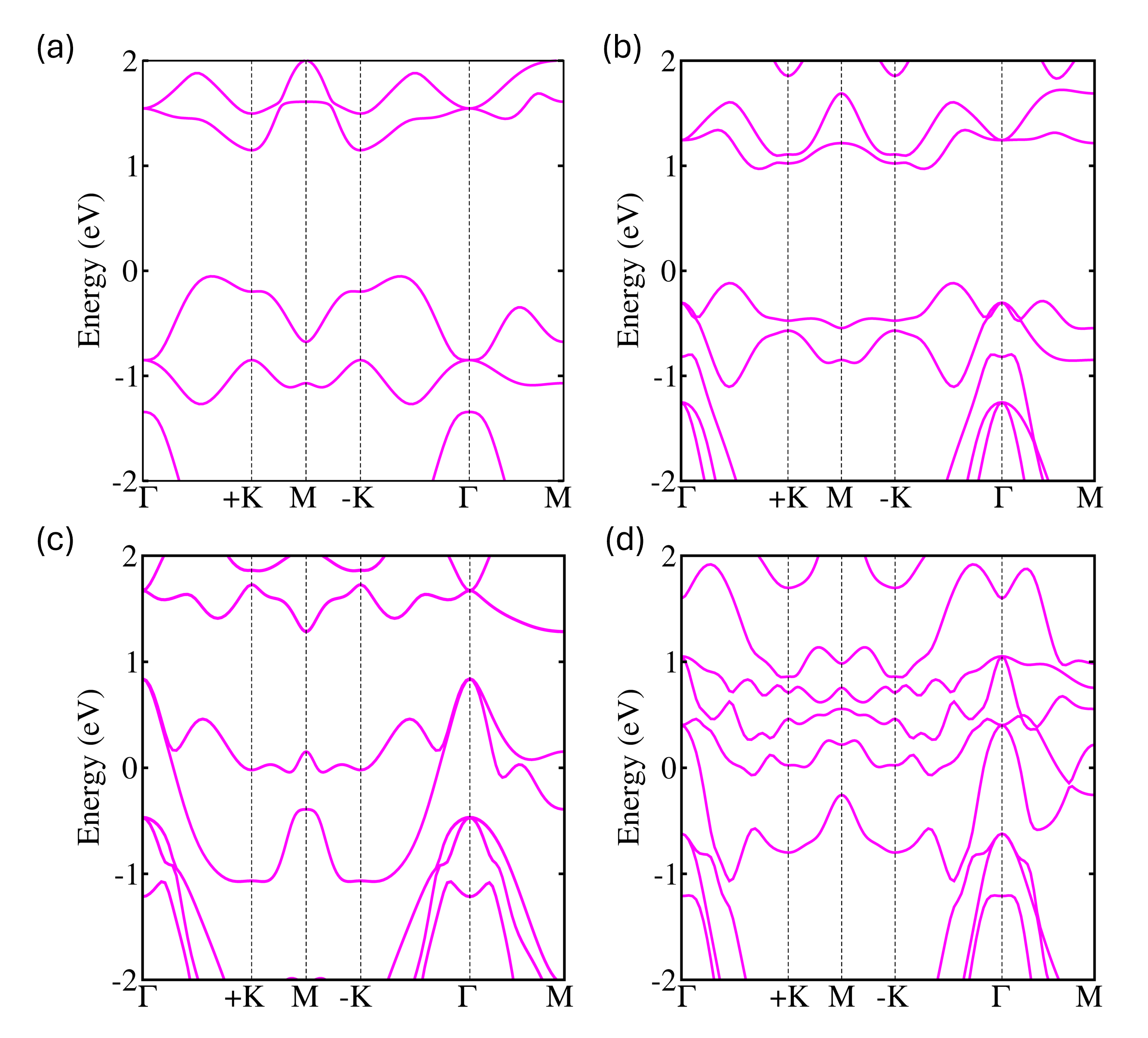}
	\caption{The electronic band structure of (a) Fe$_2$CF$_2$ showing well defined valleys at $+$K and $-$K points in the conduction band, (b) Fe$_2$CCl$_2$, (c) Fe$_2$CO$_2$, and (d) Fe$_2$CS$_2$ showing that there are no well defined valleys at high symmetry points.}
	\label{pnew}
\end{figure*}

\begin{figure*}[htp]
	\includegraphics[width=0.3\textwidth]{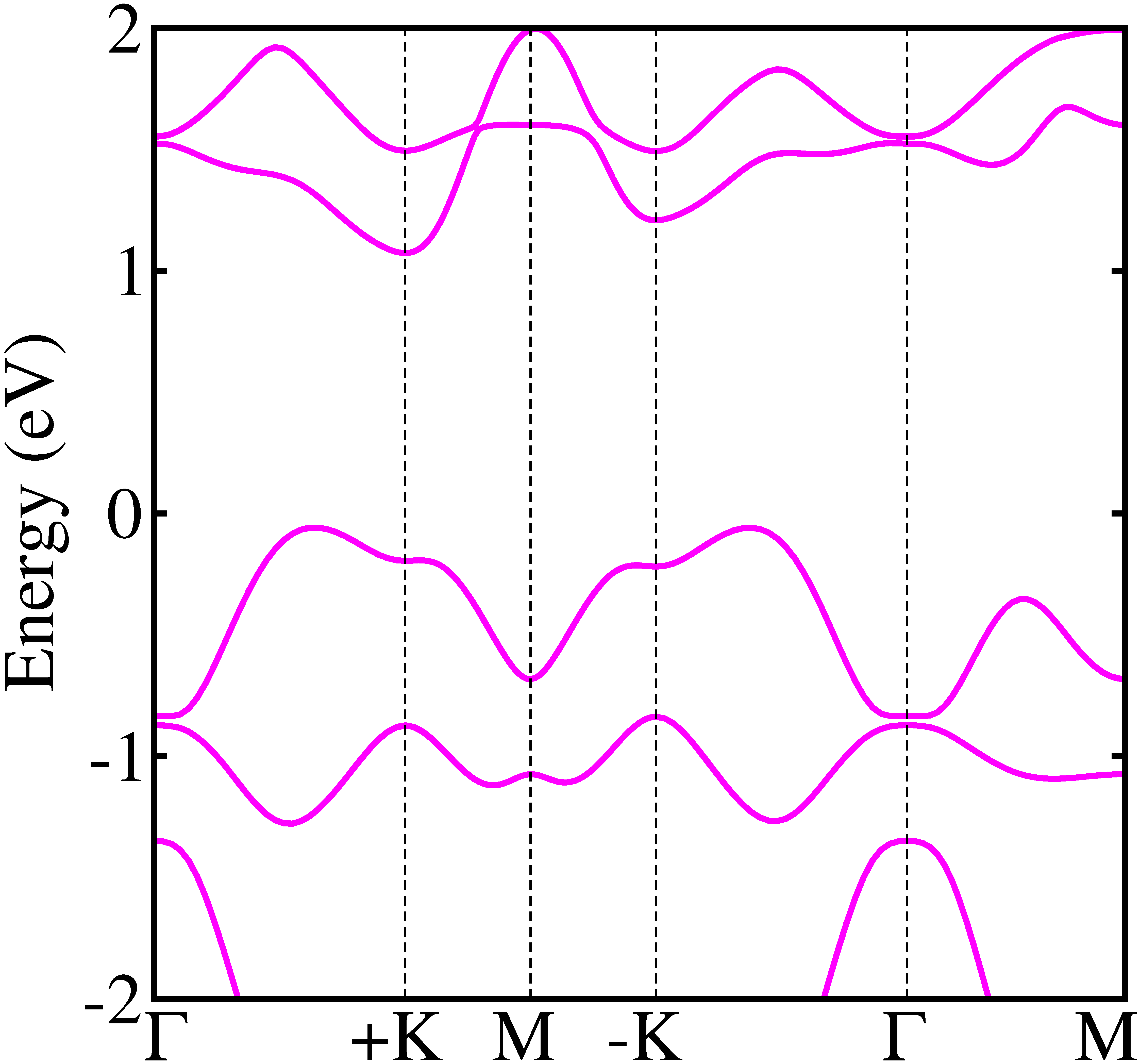}
	\caption{The electronic band structure of Fe$_2$CF$_2$ MXene with SOC showing valley polarization of magnitude 100 meV.}
	\label{pnew2}
\end{figure*}
\newpage

\section{O\MakeLowercase{rbital projected band structure of} F\MakeLowercase{e}$_2$C(OH)$_2$ MX\MakeLowercase{ene}.}

\begin{figure*}[htp]
	\includegraphics[width=0.7\textwidth]{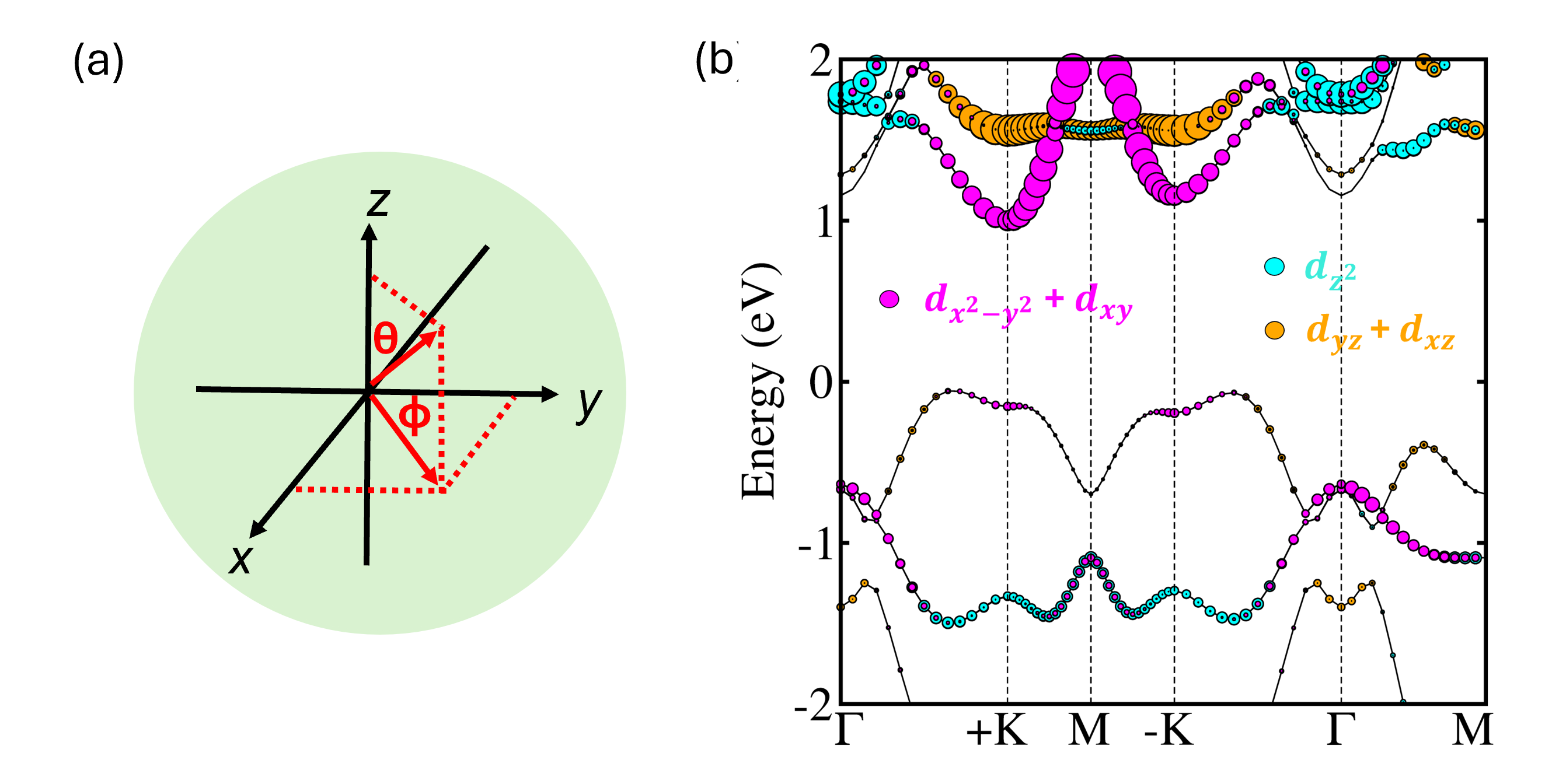}
	\caption{(a) Representation of polar angles. (b) Orbital projected band structure of Fe$_2$C(OH)$_2$ in the presence of SOC.}
	\label{p2}
\end{figure*}
\newpage

\section{T\MakeLowercase{he energy band structures for} F\MakeLowercase{e}$_2$C(OH)$_2$ \MakeLowercase{at various representative electric fields.}}
\begin{figure*}[htp]
	\includegraphics[width=1.05\textwidth]{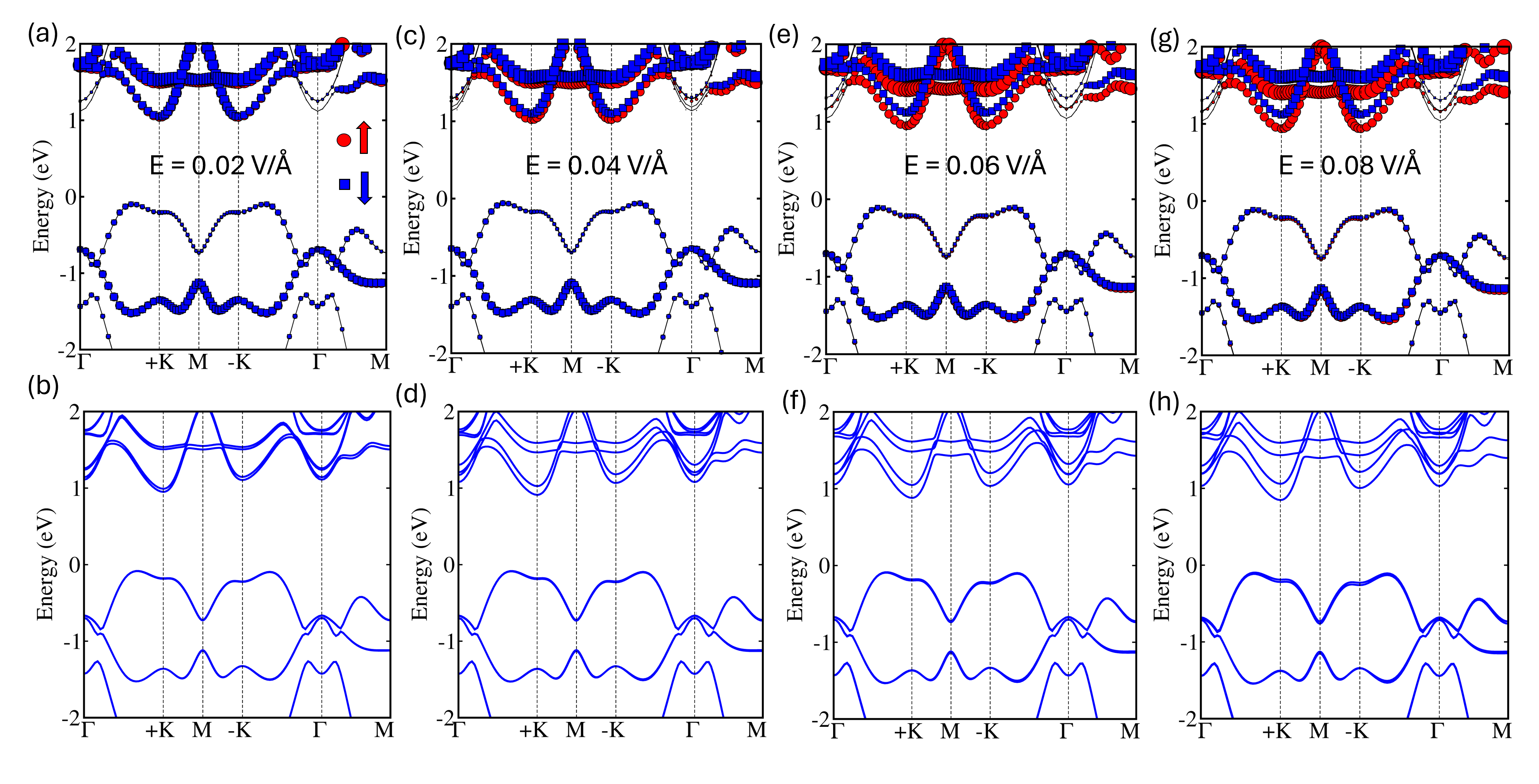}
	\caption{Band structure of Fe$_2$C(OH)$_2$ under an electric field (a) E = 0.02 V/\AA, (c) E = 0.04 V/\AA, (e) E = 0.06 V/\AA, (g) E = 0.08 V/\AA\ in the absence of SOC, and (b), (d) (e), and (h) at corresponding electric fields in the presence of SOC.}
	\label{p3}
\end{figure*}
\newpage

\section{R\MakeLowercase{eversal of the spin order of spin splitting on reversing the direction of electric field}.}
\begin{figure*}[htp]
	\includegraphics[width=0.3\textwidth]{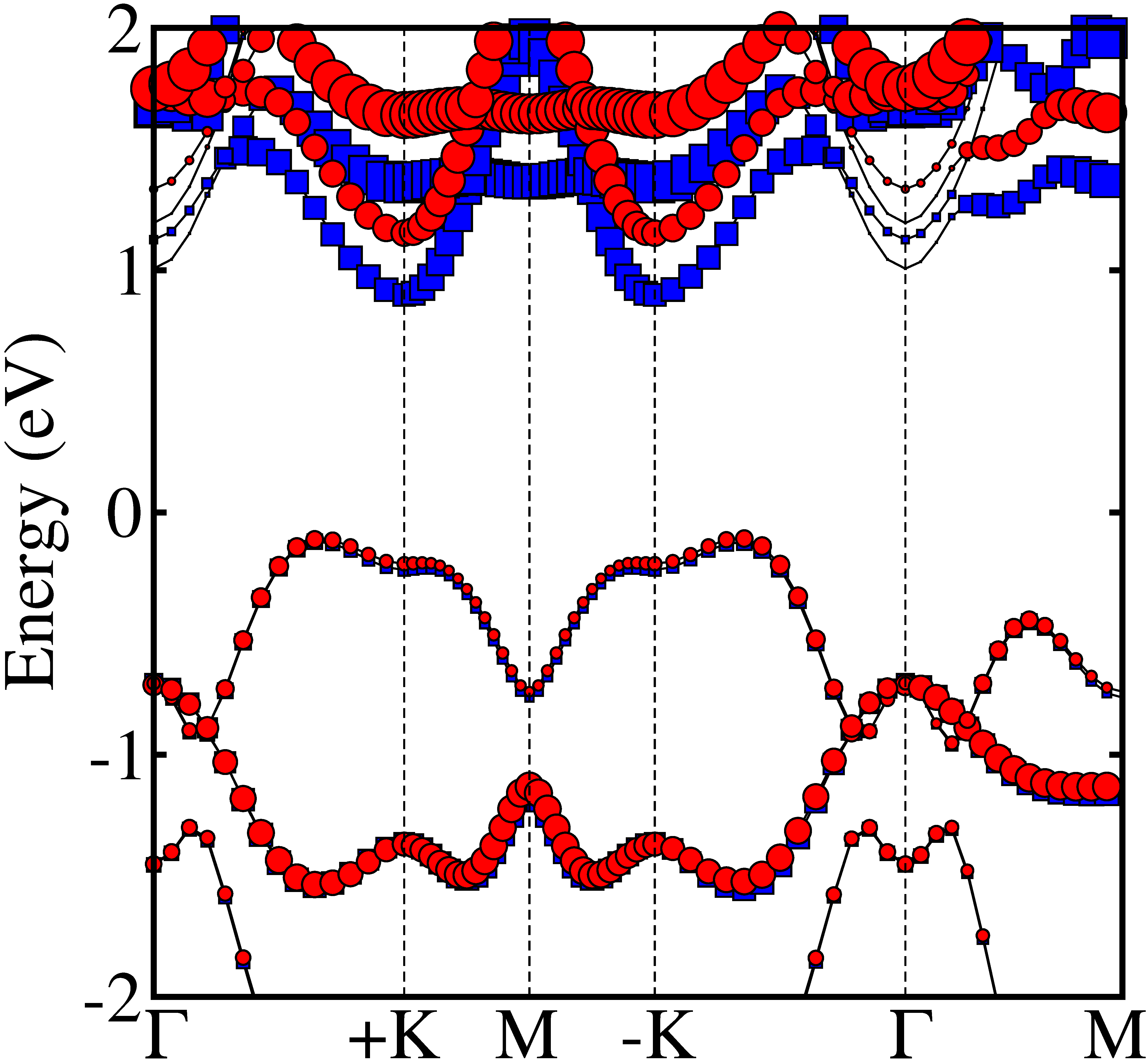}
	\caption{Reversal of the spin order of spin splitting on reversing the direction of electric field.}
	\label{p4}
\end{figure*}
\newpage

\section{D\MakeLowercase{istribution of the} B\MakeLowercase{erry curvature for} J\MakeLowercase{anus} F\MakeLowercase{e}$_2$C(OH)$_2$.}
\begin{figure*}[htp]
	\includegraphics[width=0.65\textwidth]{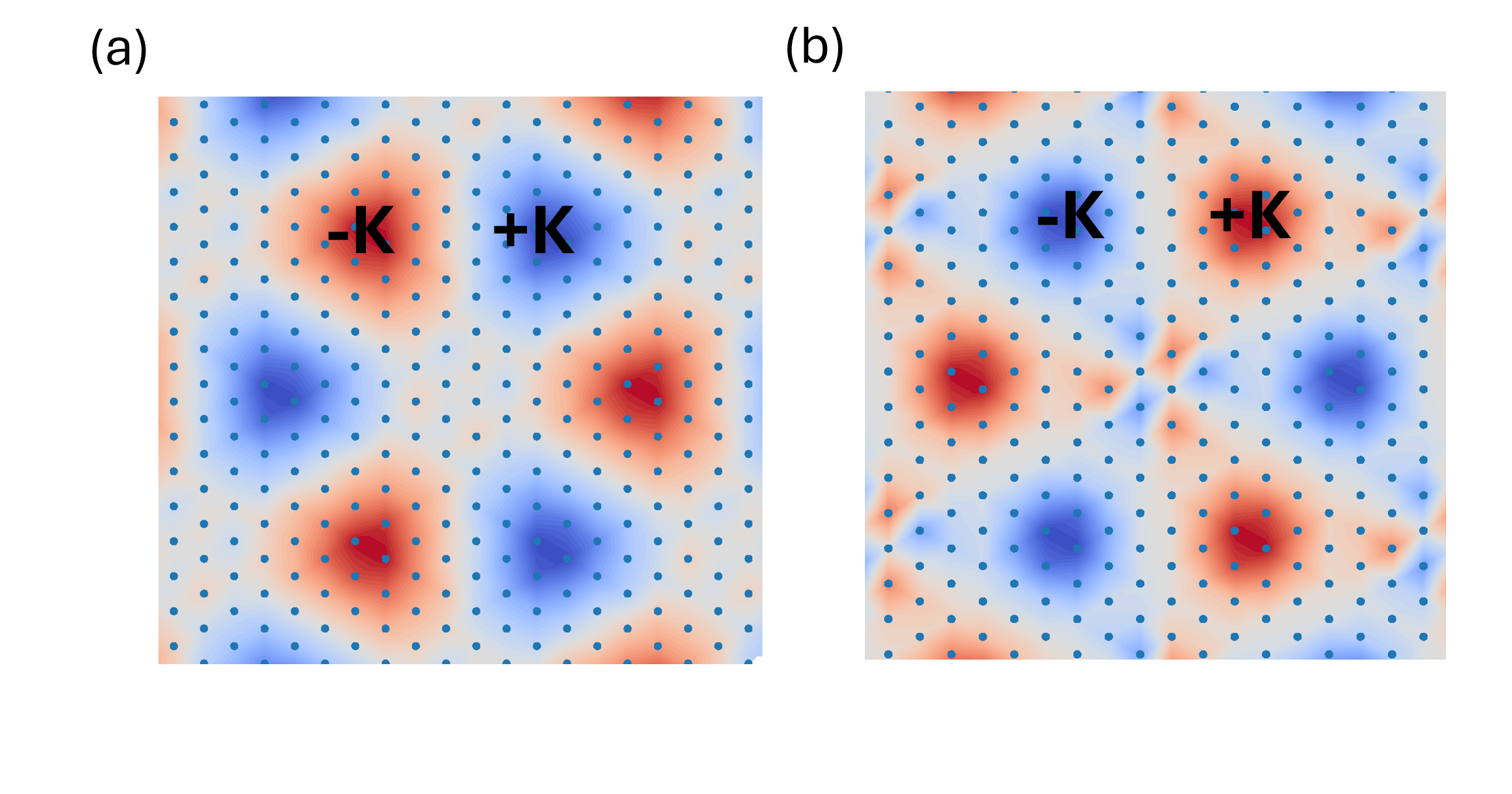}
	\caption{The distribution of (a) spin-up, and (b) spin-down Berry curvatures for Fe$_2$C(OH)F.}
	\label{p5}
\end{figure*}
\newpage

\end{flushleft}